\documentclass[cits]{PoS}
\usepackage{xspace}
\let\jnl=\rmfamily
\def\refe@jnl#1{{\jnl#1}}%
\newcommand\aj{\refe@jnl{AJ}}%
\newcommand\actaa{\refe@jnl{Acta Astron.}}%
\newcommand\araa{\refe@jnl{ARA\&A}}%
\newcommand\apj{\refe@jnl{ApJ}}%
\newcommand\apjl{\refe@jnl{ApJ}}%
\newcommand\apjs{\refe@jnl{ApJS}}%
\newcommand\ao{\refe@jnl{Appl.~Opt.}}%
\newcommand\apss{\refe@jnl{Ap\&SS}}%
\newcommand\aap{\refe@jnl{A\&A}}%
\newcommand\aapr{\refe@jnl{A\&A~Rev.}}%
\newcommand\aaps{\refe@jnl{A\&AS}}%
\newcommand\azh{\refe@jnl{AZh}}%
\newcommand\memras{\refe@jnl{MmRAS}}%
\newcommand\mnras{\refe@jnl{MNRAS}}%
\newcommand\na{\refe@jnl{New A}}%
\newcommand\nar{\refe@jnl{New A Rev.}}%
\newcommand\pra{\refe@jnl{Phys.~Rev.~A}}%
\newcommand\prb{\refe@jnl{Phys.~Rev.~B}}%
\newcommand\prc{\refe@jnl{Phys.~Rev.~C}}%
\newcommand\prd{\refe@jnl{Phys.~Rev.~D}}%
\newcommand\pre{\refe@jnl{Phys.~Rev.~E}}%
\newcommand\prl{\refe@jnl{Phys.~Rev.~Lett.}}%
\newcommand\pasa{\refe@jnl{PASA}}%
\newcommand\pasp{\refe@jnl{PASP}}%
\newcommand\pasj{\refe@jnl{PASJ}}%
\newcommand\skytel{\refe@jnl{S\&T}}%
\newcommand\solphys{\refe@jnl{Sol.~Phys.}}%
\newcommand\sovast{\refe@jnl{Soviet~Ast.}}%
\newcommand\ssr{\refe@jnl{Space~Sci.~Rev.}}%
\newcommand\nat{\refe@jnl{Nature}}%
\newcommand\iaucirc{\refe@jnl{IAU~Circ.}}%
\newcommand\aplett{\refe@jnl{Astrophys.~Lett.}}%
\newcommand\apspr{\refe@jnl{Astrophys.~Space~Phys.~Res.}}%
\newcommand\nphysa{\refe@jnl{Nucl.~Phys.~A}}%
\newcommand\physrep{\refe@jnl{Phys.~Rep.}}%
\newcommand\procspie{\refe@jnl{Proc.~SPIE}}%
\title{GRIPS and the Perspective of \\ Next-generation Gamma-ray Surveys}

\ShortTitle{GRIPS}

\author{\speaker{Roland Diehl}\\      
 and
the GRIPS collaboration\thanks{www.grips-mission.eu}\\% \thanks{A footnote may follow.}\\
        Max Planck Institut f\"ur extraterrestrische Physik, D-85748 Garching, Germany\\
        E-mail: \email{rod@mpe.mpg.de}}

\abstract{GRIPS is one example of next generation telescopes proposed for astronomy the energy range between hard X-ray mirror instruments such as NuStar and the Fermi telescope. The Compton telescope principle is an advantageous concept in view of background suppression, imaging sensitivity within a large field of view and energy range, and capability to measure polarization. The diversity of astrophysical sources at high energies (diffuse emission from cosmic-ray interactions, nuclear lines from point-like and diffuse sources, accreting binaries, cosmic-ray acceleration sites, novae and supernovae, GRBs) presents a challenge, and in particular emphasizes the need for large fields of view and surveys. We  discuss the astrophysical challenges which are expected to remain after the extended INTEGRAL mission, and how such a next-generation survey at low-energy gamma-rays would impact on these. We argue that qualitatively new and more direct insights could be obtained on cosmic high-energy phenomena and their underlying physical processes. }

\FullConference{8$^{th}$ INTEGRAL Science Workshop\\
		{\it The Restless Gamma-Ray Universe} \\
		 October 27-30, 2010\\
		 Dublin, Ireland}

\begin{document}

\section{Science Issues in High-Energy Astrophysics} %%%%%%%%%%%%%%%%%%%%%%%%%%%%%%%%%%%
The field of high-energy astrophysics is very broad, both in its observational windows and in its astrophysical diversity. Often, {\it high} energies are understood as {\it high temperatures}, meaning the hottest cosmic objects radiating in thermal emission, i.e. cosmic plasma radiating at X-ray energies. Candidate sources are accreting compact objects, from white dwarfs through neutron stars and black holes from stellar to supermassive, and gamma-ray bursters. One may, alternatively, associate  {\it high-energy source processes} with the field of high-energy astrophysics, which then encompasses {\it non-thermal} processes such as radioactive decay and nuclear de-excitation, Bremsstrahlung, synchrotron and inverse-Compton radiation, and pion decay. In this case, the energy bands of interest cover about 8--10 orders of magnitude on the frequency scale, and hence a diversity of ground- and space-based telescopes. The {\it astrophysical issues} discussed in the context of high-energy astrophysics comprise the {\it origins of cosmic rays} and {\it relativistic particle acceleration}, the {\it physics in extreme gravitational or magnetic fields (including accretion onto compact objects, and pulsars and magnetars}, the astrophysics of {\it cosmic explosions (supernovae, novae, gamma-ray bursts, and type-I X-ray bursters)}, and {\it cosmic nucleosynthesis sources}. 

\noindent These astrophysical question remain in the focus of science: \hfill\break
%\begin{itemize}
%\item{}
\indent -- Which are the Cosmic Sources of non-thermal Emission?\hfill\break
%\item{}
\indent -- How did Massive Stars influence the Early Universe and Galaxies?\hfill\break
%\item{}
\indent -- What is the Role of SMBHs in Galaxies (Blazars)?\hfill\break
%\item{}
\indent -- How does Matter flow onto Compact Objects, and make Jets?\hfill\break
%\item{}
\indent -- How do Supernova Explosions work?\hfill\break
%\item{}
\indent -- How are Cosmic-Ray Particles accelerated?\hfill\break
%\item{}
\indent -- How do Pulsars and Magnetars work?\hfill\break
%\item{}
\indent -- What can we learn from the Sun on Particle Acceleration?\hfill\break
%\item{}
\indent -- How do Positrons propagate and annihilate?\hfill\break
%\item{}
\indent -- Is there a HE Signature from Dark Matter? 
%\end{itemize}

\noindent
Considering present and planned astronomical facilities for high-energy astrophysics, we can see that the study of following cosmic objects is well underway: Accreting compact objects, cosmic relativistic particle acceleration in supernova remnants, accreting binaries, pulsars, and active galaxies. Less well covered still are the less-efficient radiation from more diluted cosmic plasma in hot interstellar and intergalactic medium, and of radiation signatures related to atomic nuclei.

\section{High-Energy Astrophysics Experiments} %%%%%%%%%%%%%%%%%%%%%%%%%%%%%%%%%%%

INTEGRAL \cite{2003A&A...411L...1W} contributed in its way to this field through data at hard-X and low-energy gamma-rays. Major astrophysical findings were the deeply-embedded sources (mostly high-mass X-ray binaries, i.e. young and massive stars), the high-energy emission tails from extremely-magnetized neutron stars, and the puzzling morphology of positron annihilation emission throughout the Galaxy. New insights on what might be useful astronomical challenges for astrophysics insights became evident from the polarization signatures in the Crab and a gamma-ray burst (GRB041219a), from the new source types such as embedded high-mass binaries, the magnetized neutron stars,  and the binaries interacting with their surroundings, all of which only became visible at energies above the 10-keV regime. More insights into nuclear emission was obtained from the spectroscopic precision in the bright $^{26}$Al line and from detection of $^{60}$Fe, but the hoped-for detections of the plausibly-expected nuclear emission in the MeV range from high-energy collisions could not be obtained. Thus, gamma-ray line spectroscopy could provide interesting and new insights on massive-star interiors through the $^{26}$Al and $^{60}$Fe measurements. But neither the promises from nova lines ($^9$Be and $^{22}$Na) nor from supernova lines ($^{56,57}$Ni and $^{44}$Ti) with their potential to clarify these cosmic explosions from measurements related to their interiors could really be fulfilled. Nuclear de-excitation lines (at 2.2~MeV from neutron captures, and 4.4 or 6.1~MeV from $^{12}$C and $^{16}$O de-excitation were seen, expected to be smoking guns of low-energy cosmic ray interactions in interstellar space or near compact objects. INTEGRAL was not sensitive enough to see at least a small sample of such sources.   

ESA's {\it Cosmic Vision} program provides a framework to gather ideas on promising space-based experiments for further insights, with above high-energy astrophysics themes included explicitly in its scope and list of themes: {\it Theme 3} of   {\it Cosmic Vision}  is called {\it the evolving high-energy universe}, and includes the themes discussed in the previous Section, although {\it $\gamma$-rays} are explicitly mentioned only in the context of black holes, surprisingly.

\section{The GRIPS Perspective} %%%%%%%%%%%%%%%%%%%%%%%%%%%%%%%%%%%
Among the diversity of concepts pursued to measure cosmic photons at energies above the regime of focusing optics (i.e. X-rays up to possibly 100~keV) and below the regime of ground-based Cerenkov or air shower instruments, the tracking chamber concept appears most promising in collecting photon interaction events both efficiently and also including imaging and background-rejection information in the measurement. At the low-energy end, interactions of photons with detector material occur mostly through Compton scattering (up to several MeV), while towards higher energies pair creation becomes the dominant interaction process. The main instrument for the GRIPS mission proposed for ESA's {\it Cosmic Vision} program and described below \cite{2008ExA...tmp...23G}, therefore, is a {\it Compton telescope} with a Si tracker as its central component, to provide secondary-particle identification and tracking for both these photon interaction types. 

\begin{figure}[th] %%%%%%%%%%%%%%%%%%%%%%%%%%%%%%%%%%
\centering
% \vspace{-50pt}\hspace{-0.7cm}
   \includegraphics[width=0.7\textwidth]{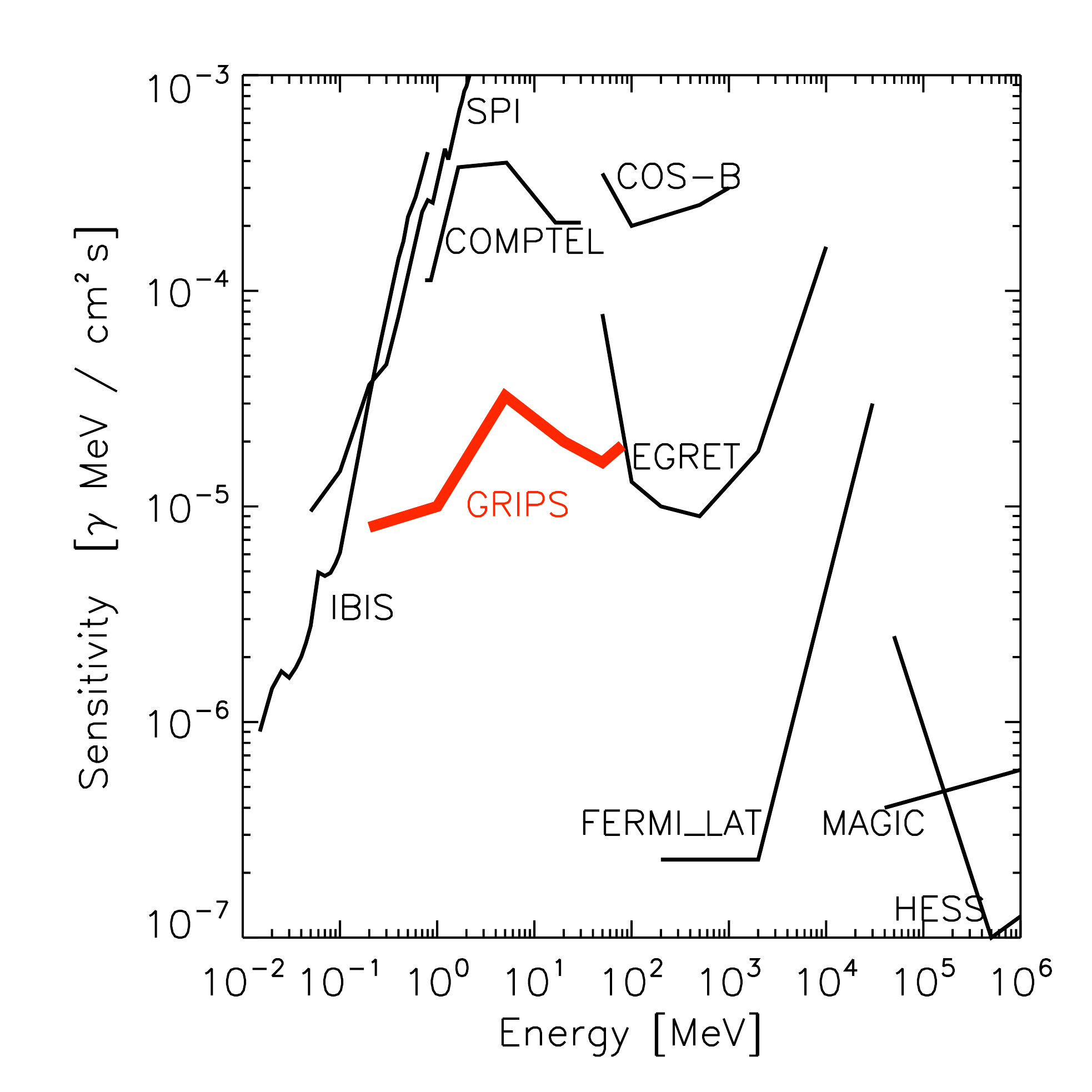}
\caption{GRIPS will allow a major sensitivity improvement
in an energy range (between hard X-rays and GeV $\gamma$-rays) which 
has been poorly explored, yet holds unique information for a wide range 
of astrophysical questions. The curves are for an exposure of 10$^6$ sec,
$\Delta E = E$, and an $E^{-2}$ spectrum.}
\label{instrument_sens.fig}
\end{figure}%%%%%%%%%%%%%%%%%%%%%%%%%%%%%%%%%%

The {\it MeV-gap} in current instrument sensitivity encompasses the photon energy range between hard X-rays of 0.2\,MeV and$\gamma$-rays of 
80\,MeV (Fig.~\ref{instrument_sens.fig}). The GRIPS mission \cite{gik09} is designed to improve the  sensitivity here by a factor of 40 compared to previous missions.
Therefore, the GRIPS all-sky survey with $\gamma$-ray imaging,  polarimetry, and spectroscopy holds a high promise of new discoveries and of precision diagnostics of primary high-energy processes.

\subsection{Instrumentation and Mission}

The GRIPS concept combines a Compton and pair telescope based on the latest 
developments in nuclear and high-energy physics laboratories (Fig. \ref{3D}). 
Modern 3D position-sensitive and space-proven detectors with advanced 
(nanosecond level) readout technology will ensure unprecedented background 
rejection capability to guarantee the above sensitivity leap. Taking 
advantage of the Compton-scattering physics, GRIPS will also be a very 
sensitive polarimeter.  
The energy resolution of 3\% at photon energies of 
1 MeV renders the gamma-ray telescope ideal for the study of broadened emission
lines from cosmic explosions such as supernovae. Moreover,  the extent of such 
spectroscopic performance throughout the entire nuclear energy range should finally enable 
pioneering astrophysical studies of nuclear de-excitation and nuclear
resonance absorption lines, and thus enable nuclear line spectroscopy similar to the rich field of atomic line spectroscopy, thus expanding beyond the radioactivity lines studied up to now.  
The limitation in imaging resolution which is intrinsic to the detection 
physics in the MeV band can be compensated by detecting secondary emission 
from the same sources with the foreseen auxiliary X-ray and NIR telescopes with their 
sub-arcmin angular resolutions.

\begin{figure}[th]%%%%%%%%%%%%%%%%%%%%%%%%%%%%%%%%%%
%\hspace{-0.2cm}
\includegraphics[width=0.5\columnwidth]{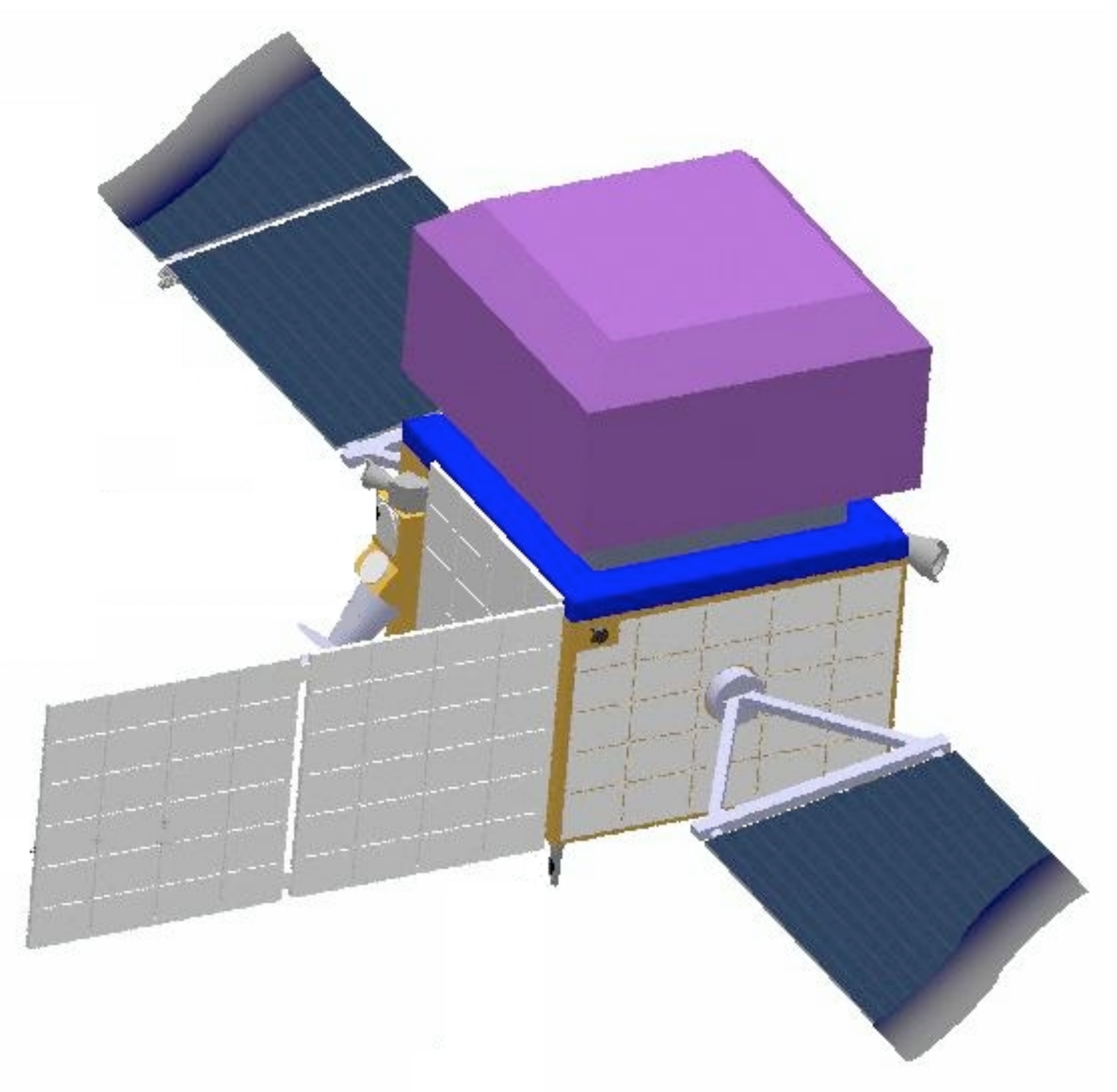}
\hfill
\includegraphics[width=0.48\columnwidth]{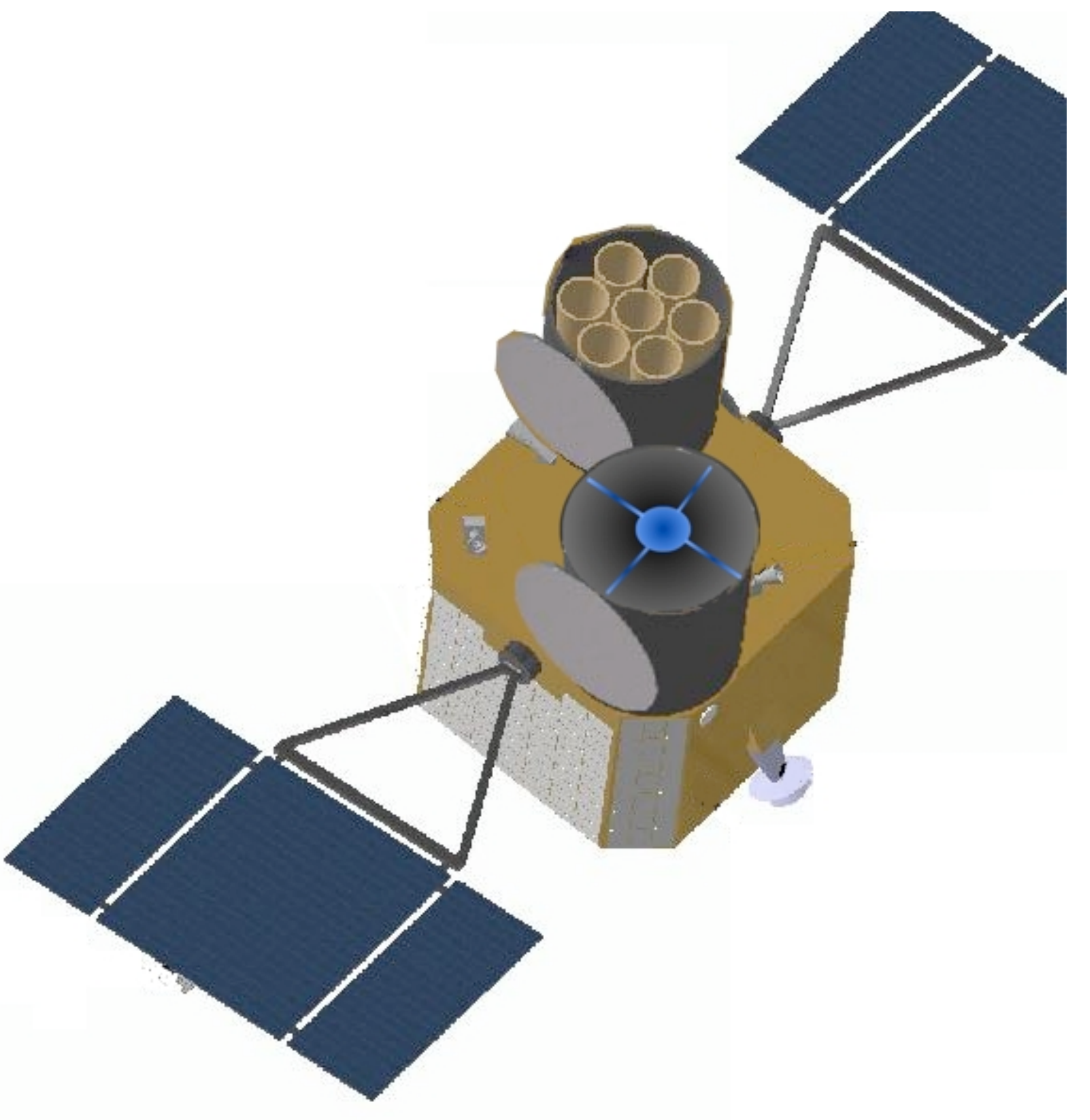}
%\includegraphics[width=0.45\columnwidth]{greiner_f3b.ps}
%\hspace{.2cm}
\caption[GRIPS_2-sat]{GRIPS configuration in the two-satellite option, 
where the GRM is on one satellite (left), and XRM and IRT on the other
(right). The GRM satellite would just do the zenith scanning all-sky survey,
while XRM/IRT would re-point with the whole (second) satellite to GRBs,
similar to {\it Swift}. }
\label{fig_2sat}
\end{figure}

GRIPS should be launched a Soyuz-Fregate rocket launched from Kourou for a projected lifetime of 10 years into a low-altitude,
equatorial orbit (LEO) to minimize the background. 
GRIPS consists of two satellites,
flying in a {\it close-pair} configuration (Fig. \ref{fig_2sat}): 
one satellite with the main instrument, the 
Gamma-Ray Monitor (GRM) including 500 kg 
of LaBr$_3$ scintillator crystals, the other with auxilliary instruments, an X-Ray Monitor XRM and an
Infra-Red Telescope (IRT). Both satellites should be 3-axis stabilized.
The gamma-ray telescope/satellite of GRIPS is planned to be continuously 
pointing at the zenith, thus monitoring ca. 80\% of the sky over 
each orbit for transient events, including gamma-ray bursts. 
The X-ray/infrared telescope satellite should have the capability of 
autonomously slewing for follow-up observations of gamma-ray bursts. 
As an added value, positions and fluxes of
transient alerts can be provided to the community. Gamma-ray bursts
in the redshift range $7<z<35$ would be recognized within minutes by IRT,
and report in near-real-time.

%%%%%%%%%%%%%%%%%%%%%%%%%%%%%%%%%%

\subsection{Prospected Science Results}

{\bf Gamma-Ray Bursts and First Stars:}
Unrivaled by any other method, the detection of highly penetrating 
$\gamma$-rays from cosmological $\gamma$-ray bursts 
shed  light on the first massive stars and galaxies which formed
during the dark ages of the early Universe.
With its energy coverage up to 80 MeV, GRIPS could firmly establish 
the high energy component seen in addition to the canonical
Band funtion in one {\it CGRO}/EGRET ($>$10 MeV) and  
one {\it Fermi}/LAT burst ($>$100\,MeV) in  much  larger numbers,  
and characterize its origin through polarisation signatures. 
GRIPS could measure the degree of polarisation of the
prompt $\gamma$-ray burst emission to a few percent accuracy for more
than 10\% of the detected GRBs, and securely
measure how the degree of polarisation varies with energy
and/or time over the full burst duration for dozens of bright GRBs.
Also, the delay of GeV photons relative to emission at $\sim$ hundred keV,
observed in a few GRBs with {\it Fermi}/LAT, manifests itself already
at MeV energies in {\it Fermi}/GBM, and would thus be a science target
for GRIPS. These observations could thus enable a clear identification of the prompt 
GRB emission processes, and determine the role played by magnetic fields.

GRIPS would be expected to detect about 650 GRBs yr$^{-1}$, a large fraction of of these 
at high-redshift ($\sim$30 GRBs yr$^{-1}$ at $z>5$, and $\sim$22 GRBs
at $z>10$). The  7-channel
near-infrared telescope (IRT) would improve  the localization to the 
required arcsecond
level and  determine  photometric redshifts for  the bulk  of the
most distance ($z>7$) sources. 
It would allow to measure the incidence of gas and metals through X-ray 
absorption
spectroscopy and line-of-sight properties by enabling NIR spectroscopy
with {\it JWST}. 
 
If  the  GRB  environments contain total hydrogen column densities of 
10$^{25}$\,cm$^{-2}$, or higher, GRIPS holds the promise of measuring  
redshifts directly from the $\gamma$-ray spectrum via nuclear resonances, 
and will be sensitive to do so beyond z$\sim$13.

GRIPS would also detect a handful of short GRBs at $z < 0.1$, enabling a
potential discovery of correlated gravitational-wave and/or neutrino signal.

{\bf Blazars:}
GRIPS would catalogue about 2000 blazars, probing blazar evolution to large
redshifts. Such observations pinpoint the most massive halos at large
redshifts, thus could severely constrain models of structure evolution.
Such a large sample could establish their (evolving) luminosity
function and thus determine the fractional contribution of blazars to the
diffuse extragalactic background. GRIPS is expected to detect $\sim$10
blazars at z $>$ 8.
Studies of the nonthermal radiation mechanisms would be supported
through spectro-polarimetric measurements. The link between the inner 
accretion disk and the jet could be probed with correlated variability 
from the thermal to the nonthermal regime, using GRIPS auxiliary 
instruments. This would localize the region of high-energy emission.

%Through nuclear lines detected in nearby AGNs, and through tracing 
%variability, GRIPS will probe the injection of accelerated particles 
%into the jet plasma.

{\bf Supernovae and Nucleosynthesis:}
The primary energy source of supernova (SN) light is radioactive decay. 
The first direct measurement of the Ni decay chain gamma-rays in Type~Ia SNe 
would provide key insight into their explosion physics and
disentangle progenitor channels. This would place the luminosity 
calibration of Type~Ia SNe on physical rather than empirical footings, for these events that are considered as standard candles in cosmology.
The otherwise unobtainable direct measurement of the inner ejecta and the 
explosive nucleosynthesis of core
collapse supernovae through $^{56}$Ni/$^{44}$Ti would allow to establish a 
physical model for these important terminal stages of massive-star evolution.
Explosion asymmetries 
%\cite{2010ApJ...714.1371H} 
and the links to long GRBs are important aspects herein.
The fraction of nearby
pair-instability supernovae from very massive stars would be unambiguously 
identified through their increased brightness from copious radioactivity.
It is such observations which are crucial for complementing  envisaged
neutrino and gravitational wave measurements, and for our understanding of 
cosmic chemical evolution.

{\bf Cosmic Rays:}
Nuclear de-excitation lines of abundant isotopes like $^{12}$C and $^{16}$O, 
the hadronic fingerprints of cosmic-ray acceleration, 
are expected to be discovered with GRIPS.
Understanding the relative importance of leptonic and hadronic processes, 
and the role of cosmic rays in heating and ionizing molecular clouds 
%and thus seeding interstellar chemistry
will boost our understanding of both relativistic-particle acceleration and 
the cycle of matter.

{\bf Magnetars:}
The measurements of instabilities in the magnetospheres
of magnetars, which are expected to lead to few-hundred keV to possibly 
MeV-peaked emission, 
could provide new aspects for the field of plasma physics and magnetic field impacts therein. Note that solar flare origins are still unclear, and the extreme magnetar flares may shed light on reconnections or other readjustments in magnetic fields far from the surface of a star.

{\bf Annihilation of Positrons:} 
The riddle of positron annihilation in the Galaxy is in need for constraining the different candidate annihilation source types. 
GRIPS could search for  positron escape from candidate sources along the galactic plane
through annihilation $\gamma$-rays in their vicinity. For several microquasars
and pulsars, point-source like appearance is expected if the local annihilation
fraction $f_{local}$ exceeds 10\% (I$_{\gamma}\sim10^{-2}\cdot f_{local}$
ph~cm$^{-2}$s$^{-1}$).
With GRIPS we could cross-correlate annihilation $\gamma$-ray images with
candidate source distributions, such as $^{26}$Al and Galactic diffuse
emission above MeV energies (where it is dominated by cosmic-ray interactions
with the ISM) [both also measured with GRIPS at superior quality], 
point sources derived from {\it INTEGRAL}, {\it Swift}, {\it Fermi}, and 
{\it H.E.S.S./MAGIC/VERITAS/CTA} measurements,
%of pulsars and accreting binaries, 
and with candidate dark-matter related emission profiles.
GRIPS wouldl deepen the current INTEGRAL sky sensitivity by at least an 
order of
magnitude in flux, at similar angular resolution. Comparing Galactic-disk and
-bulge emission, limits on dark-matter produced annihilation emission could be placed, and
constrain decay
channels from neutralino annihilation in the gravitational field of our Galaxy.
GRIPS could also perform sensitive searches for $\gamma$-ray signatures of dark
matter for nearby dwarf galaxies.

{\bf Solar Flares:}
Solar flare observations would be a natural by-product of the continuous sky survey carried
out by GRIPS since the Sun passes regularly through its field of view. 
%The number
%of flares depends on the phase of the solar cycle during the GRIPS mission.
%Lightcurves in appropriate GRIPS energy bands and high statistics spectra will
%be obtained. Comparison of $\gamma$-ray flares with data from the fleet of
%solar-terrestrial satellites and ground level observatories will bear on the
%still somewhat limited understanding of the enigmatic flare processes. 
Gamma-rays
in the MeV regime provide the means to directly probe particle acceleration and
matter interactions in these magnetised, non-thermal plasmas. Polarisation
measurements are of great value for disentangling these dynamic processes.

\section{Summary} %%%%%%%%%%%%%%%%%%%%%%%%%%%%%%%%%%%
Astrophysics of high-energy processes needs to close the observational gap remaining from the relatively poor sensitivity obtained so far in the regime between hard X-rays and GeV or higher-energy gamma-rays. This range of the electromagnetic spectrum of cosmic photons holds unique messages which cannot be obtained otherwise, and also allows to measure directly several key processes which currently are claimed to be understood or explored yet depend on sometimes uncertain models or assumed parameters. GRIPS is one of the ideas to address this, and to invest thus into qualitatively new astronomy. 

{\bf Acknowledgements}
The GRIPS consortium has discussed the astrophysical issues to be addressed and has assembled all project-related design and planning material. We appreciate the constructive work of this consortium over the past years, including the preparation of mission idea proposals for ESA's Cosmic Vision scheme.

%%%%%%%%%%%%%%%%%%%%%%%%%%%%%%%%%%%%%%%%%%%%%%%%%%%%%%%%%%%%%%%%%%%%%%%%%%%%%%%%%%%%%%%%%%%%%%%%%%%%%%%%%%%%%%%%%%%%%%%%%%%%
%
%\bibliographystyle{JHEP_MDL}
\bibliographystyle{plain}
%

% \bibliography{rod-references}
% \bibliography{rod-annihilation}
%\end{thebibliography}

\end{document}